\documentclass[doc, dvipsnames]{apa7}
\usepackage[T1]{fontenc}
\sffamily

\usepackage{lipsum}
\usepackage{ragged2e}
\usepackage[american]{babel}
\usepackage{amssymb}
\usepackage{amsmath}
\usepackage{array}
\usepackage{csquotes}
\usepackage{multirow}
\usepackage{tabularray}

\usepackage{hyperref}
\hypersetup{
    colorlinks=true,
    linkcolor=blue,
    citecolor=teal, 
    filecolor=green,
    urlcolor=blue}

\usepackage[style=apa,sortcites=true,sorting=nyt,backend=biber, hyperref=true]{biblatex}
\DeclareLanguageMapping{american}{american-apa}
\title{Beyond Group Means and Into the World of Individuals: A Distributional Spotlight for Experimental Effects on Individuals}
\shorttitle{Changes of Distributions as Experimental Effects}

\authorsnames[]{Roussel Rahman}
\authorsaffiliations{{Stanford University, Stanford, CA}}

\leftheader{Rahman}

\abstract{Traditionally, experimental effects on humans are investigated at the group level. In this work, we present a distributional ``spotlight'' to investigate experimental effects at the individual level. Specifically, we estimate the effects on individuals through the changes in the probability distributions of their experimental data across conditions. We test this approach on Reaction Time (RT) data from 10 individuals in a visual search task, examining the effects of (1) information set sizes and (2) the presence or absence of a target on their processing speed. The changes in individuals' RT distributions are measured using three approaches: (i) direct measurements of distributional changes are compared against the changes captured by two established models of RT: (ii) the ex-Gaussian distribution and (iii) the Drift-Diffusion model. We find that direct measurement of distributional changes provides the clearest view of the effects on individuals and highlights the presence of two sub-groups based on the effects experienced: one that shows neither effect and the other showing only the target-presence effect. Moreover, the intra-individual changes across conditions (i.e., the experimental effects) appear much smaller than the inter-individual differences (i.e., the random effects). Generally, these results highlight the merits of going beyond group means and examining the effects on individuals, as well as the effectiveness of the distributional spotlight in such pursuits. 
}


\authornote{
   \addORCIDlink{Roussel Rahman}{0000-0003-2009-4246}

  Correspondence concerning this article should be addressed to Roussel Rahman, Department of Photon Science, SLAC National Accelerator Laboratory, Menlo Park, CA.  E-mail: roussel.rahman@gmail.com}

\begin{document}
\maketitle

Experimental effects on human behavior (i.e., the changes in behavior across experimental conditions) are traditionally investigated at the level of groups. The experimental data in each condition are averaged over a large number of individuals, and the experimental effects are estimated based on the group mean differences across conditions. While group studies serve specific research needs, taking individuality into account is essential when one-size-fits-all explanations may not exist. For example, in an experiment, participants may experience different levels or types of effects, but the individuality of effects would be obscured in the group-level averages.

This concern, whether the averages represent the individuals well, is well-documented across areas of research and requires meeting specific criteria to overcome and make individual-level conclusions based on group estimates \parencite{estes1956problem, siegler1987perils, molenaar2009new, torres2016toward, gray2017pdls, rahman_gray2020topics, rahman2022dynamics, rose2013science}. Nevertheless, focusing on group means limits our understanding of the individuality of effects, which is essential for real-world applications, such as when evaluating learning of a diverse class of students or for psychiatric diagnosis and treatment of individual humans. In such cases, understanding what ails the individual -- as opposed to what affects the group -- may help to find personalized solutions to aid them. Consequently, there is a growing trend of going beyond group means and toward models of individual behavior in several areas of psychological research, such as human information processing \parencite{heathcote1991analysis, whelan2008effective, wolfe2010reaction, mittelstadt2020beyond}, human learning and skill acquisition \parencite{heathcote2000power, siegler1987perils, rahman2021precisemeasures, rahman_gray2020topics, towne2016understanding}, and precision psychiatry \parencite{torres2016toward, fernandes2017new, tozzi2024personalized}.

Despite the importance of individual-level investigations and the promise of personalized solutions, methods for investigating individual-level effects (such as the changes with treatment or development) remain limited \parencite{torres2016toward, fernandes2017new, rahman2022dynamics, rose2013science}. In particular, simple yet reliable measures to detect fine-grained changes in individuals' data of varying types can help us in important endeavors, such as early detection of mental health disorders, evaluating treatment benefits for individuals, and modeling developmental changes with aging. In this work, we explore a distributional ``spotlight'' for the effects on individuals through the changes in the probability distributions of their experimental data. Examining two experimental effects on 10 individuals in a visual search task, we demonstrate the effectiveness of the distributional changes in capturing the fine-grained changes in individuals, as well as highlighting the large individual differences of effects -- results that underline the importance of going beyond the mean and into the world of individuals. 


\section{When to Go beyond the Mean?}
Researchers across fields and eras urged us to go beyond the mean.
The mean is the expected value of a normal distribution. This entails that if we are to bet on a normally distributed value and we get to play many times, our best bet is the mean, which guarantees the maximum average profits in the long run. A lot of naturally occurring data follows the normal distribution (from human heights, weights, and IQ scores to material properties, stock prices, and motion of particles).
In such cases, our sample means from repeated sampling from the population are expected to follow a normal distribution. 
Therefore, the means from different experimental conditions can be compared to estimate the effects through the mean differences.

But what if our data is not normally distributed? For such data, the mean is no longer the expected value. One prime example is the RT distributions of individuals in information processing tasks. Individuals' RT distributions have been generally found to be asymmetric or skewed, whereas normal distributions are symmetric. Generally, a promising direction has been moving towards individual-level models that explain whole distributions of behavioral data \parencite{wolfe2010reaction, ratcliff1978theory, ratcliff2016diffusion, mittelstadt2020beyond, heathcote1991analysis}. For example, to account for the skewness, researchers adopt different distributions -- notably, Gamma, Ex-Gaussian \footnote{also known as the Exponentially Modified Gaussian Distribution} (Ex-G), and Weibull distributions.

Before diving into distributional analyses, a relevant question is: Is non-normality a rarity or the norm? Several researchers over the past four decades confirmed the prevalence of non-normality in data from psychological experiments.
In a meta-analysis of 693 datasets, \textcite{blanca2013skewness} found that 95\% of the data violated the assumptions of normality. Previously, \textcite{micceri1989unicorn} tested 440 datasets for normality and found all of them to be significantly non-normal. In short, evidence shows that very rarely our experimental data is close to being normal, so rarely that \textcite{micceri1989unicorn} dubbed normal distributions as the ``unicorn" in psychological research.


\section{Distributional Analyses of Experimental Effects on Individuals}

Typically, the distributions are explained by multi-parameter models. For example, RT data is often modeled using three-parameter Ex-G or Gamma distributions. Sometimes, we may have a theoretical cognitive model that produces the distributions of behavioral data. The Drift-Diffusion Model \parencite[DDM;][]{ratcliff1978theory} for RTs in binary-choice tasks provides an example. The experimental effects are investigated through the parameter changes across conditions. For example, \textcite{torres2016toward} used Gamma distributions to model the speed of responses across trials and demonstrated that individuals with different disorders can be categorized by the parameters (the shape and the scale) of Gamma distributions. The DDM parameters have been used to identify cognitive processes that lead to premature responses and impaired response inhibition in patients with Schizophrenia \parencite{limongi2018knowing} and slow responses in patients with Parkinson's disease \parencite{ocallaghan2017visual}.

Please see \textcite{heathcote1991analysis, whelan2008effective, balota2011moving} for reviews on the applications of Ex-G distributions. For relatively recent reviews on the DDM, please see \textcite{voss2004interpreting, voss2013diffusion, wiecki2013hddm, shinn2020flexible}. However, as the distributions are characterized by multiple model parameters, investigating the changes can be quite complex for many individuals. Moreover, each parameter captures a specific characteristic of the distribution, but they are not independent, and there are often tradeoffs among them. Therefore, the parameters are difficult to interpret in isolation, but combining the changes in different parameters into a single measure is also not straightforward.

\section{Goals and Scope of this work}

In this work, we explore a simple and general way to examine the experimental effects on individuals. Specifically, we measure the distributional changes directly by the changes in the information contained in the RT distributions. Our goal is to highlight and magnify the changes in individuals -- therefore, we refer to this method as the "spotlight" method. 

We evaluate our distributional spotlight by comparing it against two established methods for modeling RT data: (1) Ex-G distributions and (2) the DDM. As our testbed, we examine two well-known effects on human information processing: (1) the effect of information set sizes and (2) the effect of the presence or the absence of a target in the sets. Typically, the average subject demonstrates both of these effects -- that is, RT is usually slower with increasing set size and increasing amount of irrelevant information. These results indicate that humans are generally (1) serial information processors and (2) unable to inhibit or ignore irrelevant information.

However, do all individuals experience the same effects? We examine this research question using the three distributional investigations outlined above on RT data from 10 individuals in a visual search task. In the following sections, we describe the Ex-G distributions and the DDM before the two direct measures of distributional changes we use to instrument the spotlight -- Relative Entropy (RE) and the Overlapping index (OI).

\section{Modeling RT Distributions}

\subsection{Ex-G Distribution}


The Ex-G distribution (Eqn. \ref{eqn:exg_distribution}) provides a process-level explanation of RTs using its parameters. A frequently observed cause for the skewness of RT distributions is unusually long responses in some trials, which is considered a marker of attention lapses \parencite{whelan2008effective, hervey2006reaction}. The ex-Gaussian distribution is a convolution of the exponential distributions with the Gaussian distribution. 
and is characterized by three parameters: \{$\mu$ (mu), $\sigma$ (sigma), and $\tau$ (tau)\}. $\mu$ and $\sigma$ respectively correspond to the mean and the standard deviation of the Gaussian component of the convolution representing the usually fast responses. $\tau$ is the mean of the exponential component and represents the skewness due to the slow responses during attention lapses.

\begin{align}\label{eqn:exg_distribution}
p(x; \mu, \sigma, \tau) = \frac{1}{\tau} \exp\left(\frac{1}{2\tau^2} - \frac{x - \mu}{\tau}\right) \text{erfc}\left(\frac{\frac{1}{\tau}(x - \mu) - \frac{\tau}{\sigma}}{\sqrt{2}}\right)
\end{align}

\subsection{Drift-Diffusion Model of RT}

The DDM of RT \parencite{ratcliff1978theory} is a well-established theory that explains human information processing in binary choice tasks. The DDM has successfully explained human information processing speed in a broad range of tasks for studies in cognitive psychology, neuroscience, psychiatry, and health sciences \parencite[for reviews, please see][]{ratcliff2016diffusion, voss2004interpreting,  voss2013diffusion, myers2022practical}.

This model explains the RT distributions from trial-level cognitive processes underlying individuals' decision-making. Mathematically, the model can be expressed by a Stochastic Differential Equation (SDE) representing a drift-diffusion process for a decision variable $x$ (Eqn. \ref{eqn:DDM_as_SDE}). For information processing, the drift represents moving towards one of two decision boundaries based on accumulated evidence, and the diffusion represents stochastic noise that may slow us down and sometimes lead to wrong decision boundaries. Each individual (i.e., their decision processes) is characterized by a set of four or more parameters (e.g., drift rate $a$, diffusion coefficient $D$, boundary separation, initial position for priors, and non-decision time). The total RT is considered as the sum of the durations of three stages of sequential processing: (i) encode information in stimulus, (ii) process information and make a decision, and (iii) execute motor response for the decision \parencite{sternberg1969memory, sternberg2015sequential, myers2022practical}.

  \begin{equation}\label{eqn:DDM_as_SDE}
      dx = a(x,t)dt + D(x,t)dW
  \end{equation}

  \begin{equation}\label{eqn:DDM_as_PDE}
      \frac{\partial}{\partial t} p(x,t) = \frac{\partial}{\partial x}a(x,t)\ p(x,t) + \frac{1}{2} \frac{\partial^2}{\partial x^2}D^2(x,t)\ p(x,t)
  \end{equation}

The Generalized DDM (GDDM) by \textcite{shinn2020flexible} uses an alternate representation of the SDE as a Partial Differential Equation (PDE); specifically, as the Forward-Kolmogorov Equation\footnote{Also known as the Fokker-Planck Equation} (Eqn. \ref{eqn:DDM_as_PDE}). Importantly, this PDE provides a deterministic relationship for the changes in the probability distributions, $p(x)$, of the decision variable, as opposed to a stochastic one for the variable $x$ itself. In a way, we hide away the stochasticity in the distributions and retain only the deterministic portion of the changes. 

As a (sub-)goal is to explain the differences or the changes of the distributions across experimental conditions, we believe directly measuring distributional changes provides a simple and promising way to estimate the experimental effects on individuals. In this work, we explore two measures of distributional differences to examine the effects.

\section{Measuring Difference (or Similarity) between RT Distributions}
\subsection{Relative Entropy}RE or KL Divergence \parencite{kullback1951information} is often considered to be ``the most appropriate quantity to measure distinguishability between different states'' \parencite[pp. 198]{vedral2002role}. It measures the extent to which a target distribution differs from a reference distribution. If $p(x)$ and $q(x)$ are respectively the target and reference distributions of a continuous random variable $x$, RE of $p$ with respect to $q$ is \parencite{cover2012elements}:

\begin{equation} \label{eqn:RE}
    RE (p(x)||q(x)) = \int_{X}\ p(x)\ \log_2 \left(\frac{p(x)}{q(x)}\right)\ dx
\end{equation}

The symbol $||$ denotes the relativity -- that we are estimating the difference of our target distribution $p(x)$ relative to the reference distribution $q(x)$. This relativity is important as RE is not symmetric (i.e., $RE(p||q) \neq RE(q||p)$). To make it symmetric, Kullback and Leibler averaged between the two (i.e., $RE_{symmetric}(p,q) = (RE(p||q) + RE(q||p))/2$), a solution we also follow in this work.
The minimum RE is 0, corresponding to identical target and reference distributions. However, RE does not have an upper bound, depicting that the differences between two distributions can be infinite.

\subsection{Overlapping Index}
\textcite{pastore2019measuring} developed the OI to estimate the similarity between two distributions (Eqn. \ref{eqn:OI}) by calculating the overlapping area between the distributions. Although this measure is less common than RE, the OI provides an intuitive measure of distributional similarity. Relevantly, \textcite{pastore2019measuring} used the OI to estimate the sample differences between groups to perform hypothesis tests based on Bayesian evidence.

\begin{align}\label{eqn:OI}
    OI(p(x), q(x)) =& \int_{X} min(p(x), q(x)) dx \notag \\
    =& 1 - \frac{1}{2}\int_{X} |p(x) - q(x)| dx
\end{align}

The minimum OI is 0 (corresponding to no overlap in areas under the distributions), and the maximum is 1 (exact overlap). OI is a measure of similarity, whereas RE measures differences. For ease of comparison, we convert OI (and other similarity measures used in this work) to differences by subtracting from 1 (i.e., $\mathrm{Difference = 1 - Similarity}$). With this conversion, OI equals 0 with no difference between distributions, and the OI approaches 1 as the differences go to infinity.

\section{Methods}

\subsection{Experimental Task and RT Dataset}
We use the RT dataset from Experiment 1 by \textcite{wolfe2010reaction}. In this experiment, each of the 10 participants completed 4000 trials of a visual search task. In the task, participants needed to indicate the presence or the absence of a red vertical rectangle among green vertical rectangles. They performed the task in four setsizes (3, 6, 12, and 18) and two target presence conditions (present vs. absent). Therefore, for each individual in each of the $4 \times 2=8$ experimental conditions, we have 500 RT samples to fit ex-Gaussian distributions and estimate their parameters. The dataset is openly available at \url{https://search.bwh.harvard.edu/new/data_set_files.html}. Please see \textcite{wolfe2010reaction} for details of the experimental procedure.

\subsection{Estimating Model Parameters from RT Data}
Table \ref{table:six_measures} provides lists of the model parameters and the measures of their change. The Ex-G distribution parameters were estimated using maximum likelihood estimation and a differential evolution optimizer. The DDM parameters were estimated using the pyDDM package by \textcite{shinn2020flexible}, who also use differential evolution to solve for the probability distributions in the GDDM model (Eqn. \ref{eqn:DDM_as_PDE}).

We estimate the model parameters for each of the 10 individuals. Moreover, we average each model parameter for all participants to obtain a model of the ``average" subject. The group average could be obtained in other ways (such as pooling the RT samples of all individuals in one distribution before fitting them to models). However, \textcite{rouder2004evaluation} demonstrated that averaging distribution parameters usually outperforms pooling approaches for obtaining group-level distributions.

\subsection{Six Measures of Distributional Differences}\label{sec:measures_of_pdf_change}

\begin{table}[!t]
\centering
\caption{Parameters of RT Models and their Changes}
\begin{tblr}{
  colspec={|p{3cm}|p{7cm}|p{3.5cm}|}, 
  cell{2}{1} = {r=2}{},
  cell{2}{2} = {r=2}{},
  cell{4}{1} = {r=2}{},
  cell{4}{2} = {r=2}{},
  cell{6}{1} = {r=2}{},
  cell{6}{2} = {r=2}{},
  vlines,
  hline{1-2,4,6,8} = {-}{},
  hline{3,5,7} = {3}{},
}
Model            & Parameters                             & Measures of Change\\
Any Distribution & - & Relative Entropy   \\
                      &                       & Overlapping Index   \\
Ex-G Distribution & mean ($\mu$), standard deviation ($\sigma$), \& skew ($\tau$) & Cosine Similarity   \\
                      &                       & Euclidean Distance   \\
DDM & drift ($a$), diffusion ($D$), boundary separation ($B$), \& initial position ($x_0$)  & Cosine Similarity   \\
                      &                       & Euclidean Distance   \end{tblr}
\label{table:six_measures}
\end{table}

The first two measures we use are RE and OI. Their values were calculated using Ex-G distributions in Equations \ref{eqn:RE} and \ref{eqn:OI}. A challenge to calculating these measures is that, for many distributions (including Ex-G distributions), there are no closed analytical forms (or formulas) for the integrals in the equations. This problem can be side-stepped using numerical integration, which presents a different challenge: dealing with improper integrals that spread to infinity. We overcome this challenge by converting them to proper integrals with finite limits. As our functions are all Probability Density Functions (PDFs) with the area under their curves equaling 1, we use the limits corresponding to Percent Point Functions (PPFs) ranging from 0.001 to 0.999, thus accounting for 0.998 or 99.8\% of the area under the PDFs. Then, the proper integrals within finite limits are calculated using the trapezoidal method.

For the Ex-G distributions and the DDM, we combine multiple parameters into one distance or similarity metric. Viewing it as a vector geometry problem where each parameter is a dimension, we have two simple measures for the difference between two points in n-dimensional space: Euclidean Distance (ED) and Cosine Similarity (CS). ED measures the spatial distance between the points, whereas CS is the cosine of the angle between the origin vectors of the points. We use these measures for their simplicity and wide use, although they are limited in high-dimensional spaces \parencite{aggarwal2001surprising}. Finally, to aid comparisons, we convert the similarity measures in the mix (i.e., OI and CS) into dissimilarities by subtracting the similarity values from 1.

\section{Results}

\subsection{RT Distributions of Individuals}

\begin{figure*}[!t]
    \includegraphics[width=\textwidth]
    {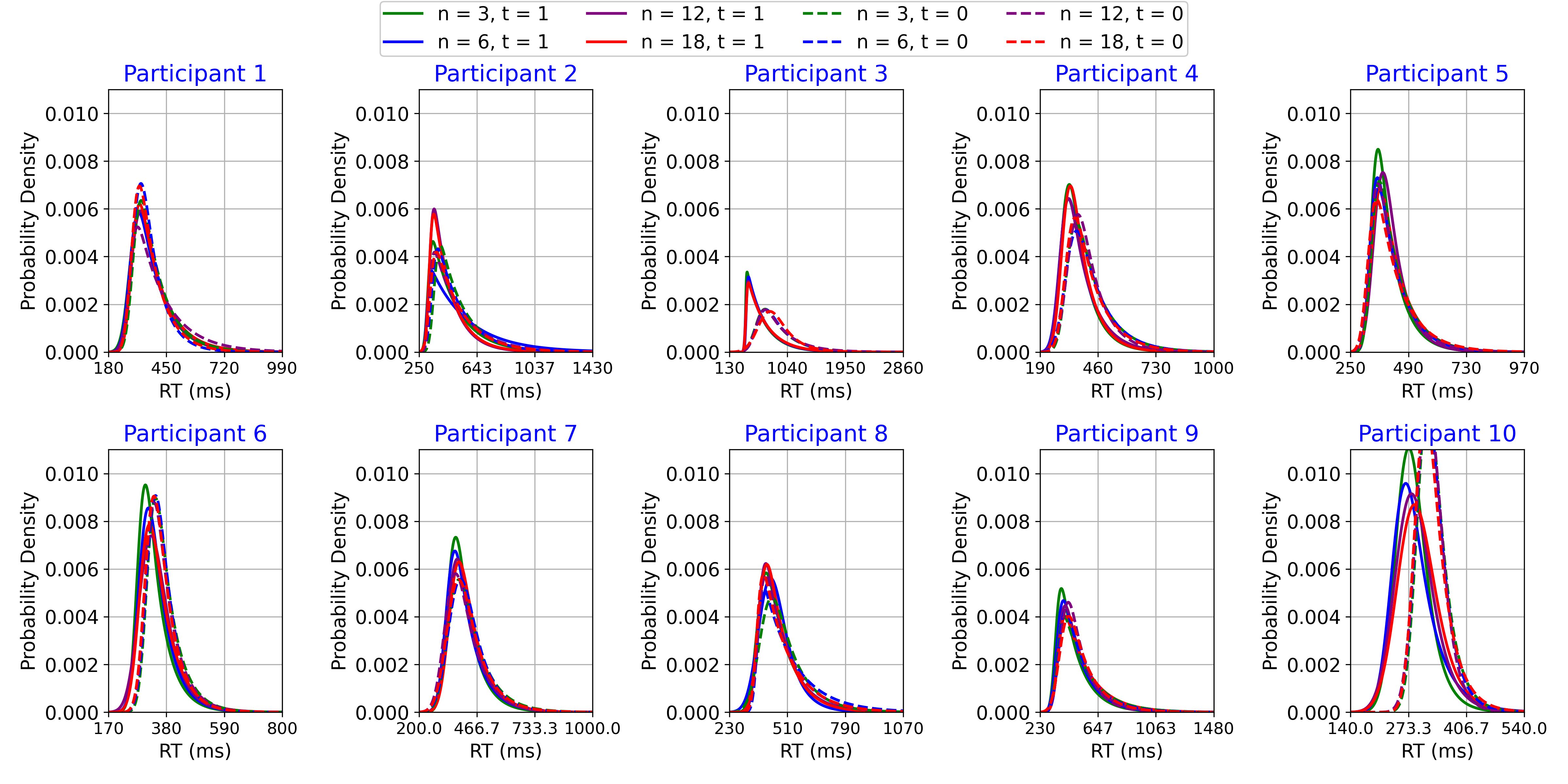}
    \caption{RT Distributions of 10 participants in all experimental conditions. The solid lines represent distributions under target-present conditions, and the dashed lines represent target-absent conditions. Each distribution is plotted in a range that contains 99.8\% of the distribution.}
    \label{fig:distributions}
\end{figure*}

\begin{figure*}[!ht]
    \includegraphics[width=\textwidth]
    {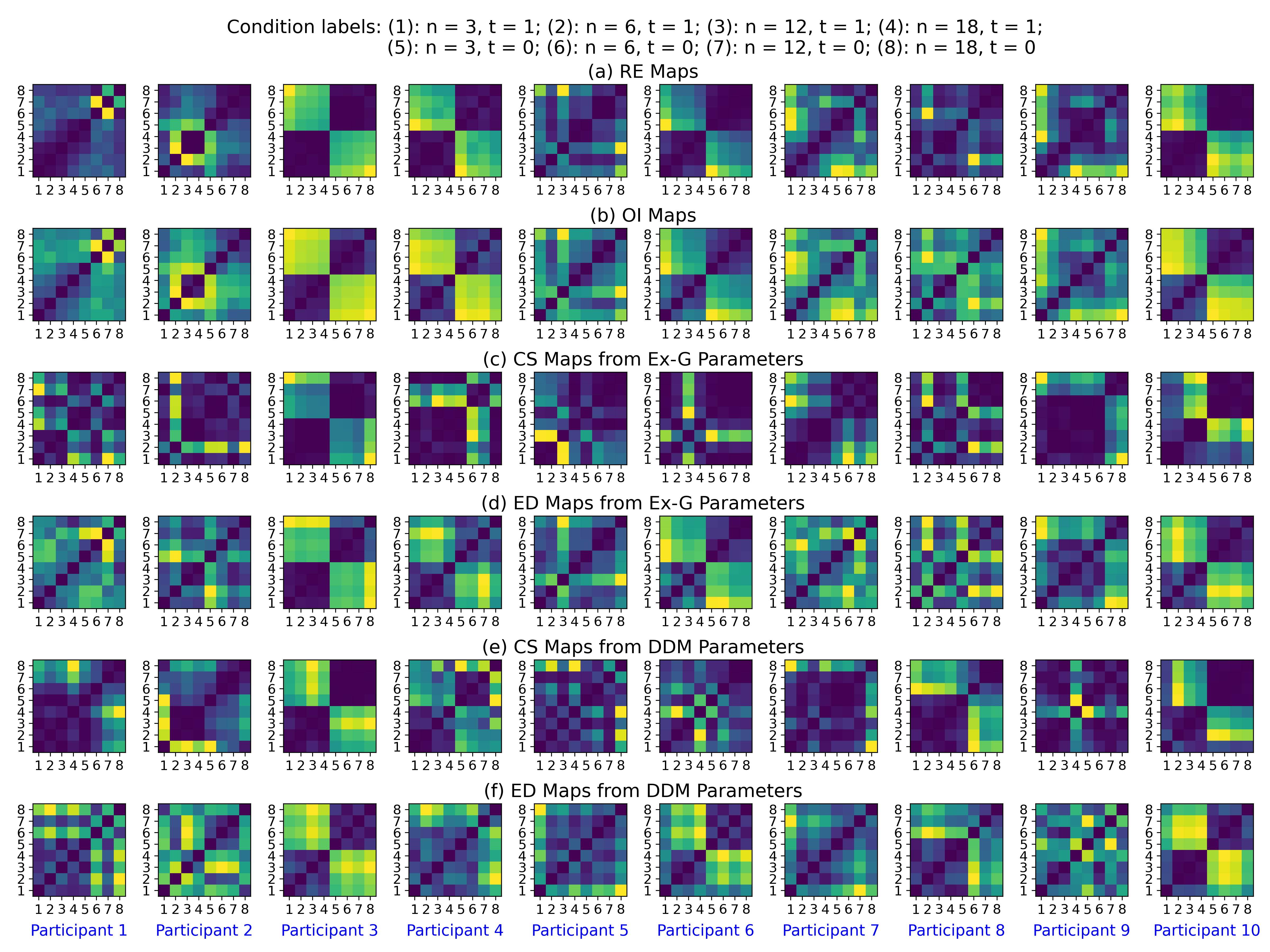}
    \caption{Distributional changes across conditions calculated using six different measures. For each measure, the distributions from all pairs of conditions are compared and presented as a map.}
    \label{fig:six_distances}
\end{figure*}

\begin{figure*}[!ht]
    \includegraphics[width=\textwidth]
    {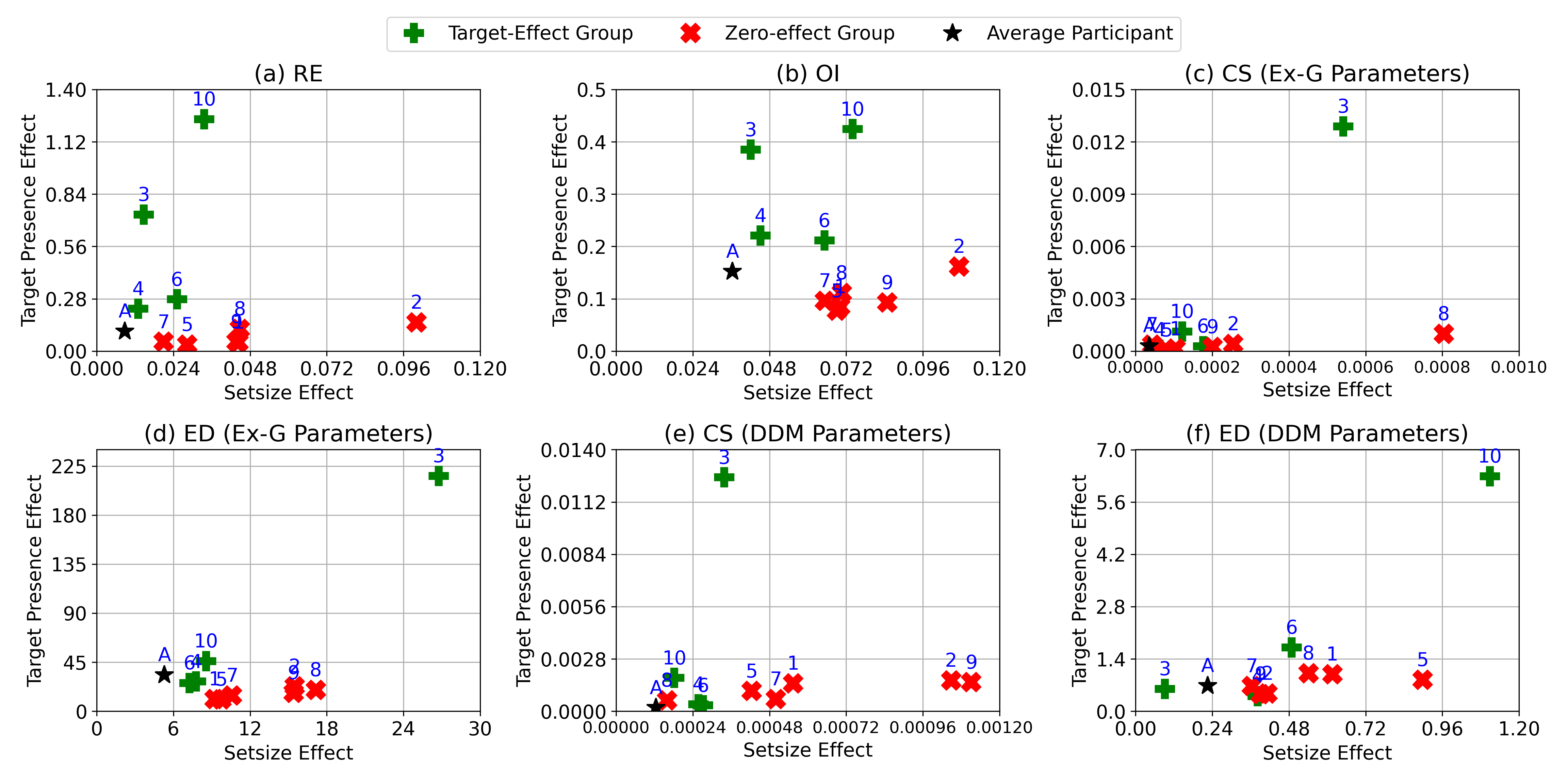}
    \caption{Average experimental effects on participants measured in six ways. As we see, RE and OI provide the clearest separation between the two groups.}
    \label{fig:effects_on_individuals}
\end{figure*}

We begin by looking at the RT distributions of each individual across all eight conditions (4 set sizes x 2 target conditions, in Fig. \ref{fig:distributions}). To capture a comparable range of each distribution without the extreme tails that spread to infinity, we fixed the plot ranges to cover 99.8\% of each distribution, using the same process described in Section \ref{sec:discussion}. 

Notably, the inter-individual differences in the distributions appear to be much higher than the intra-individual changes across conditions. For each individual, the distributions across conditions show considerable overlap and span roughly the same range on the x-axis. On the other hand, the ranges of RT values can be observed to vary drastically across individuals. These results indicate that the individuals remain more similar to themselves across conditions than to others in the group. Put another way, in this experiment, the main experimental effects on individuals (our signal) appear to be much weaker than the random effects (i.e., the noise) due to inter-individual differences within groups.

As for the changes, we observe that two dominant trends in the intra-individual changes of distributions divide the individuals into two sub-groups. The first sub-group (consisting of Participants 1, 2, 5, 7, 8, and 9) shows no to little effects of either set size or target presence, with the distributions for different conditions being nearly indistinguishable from each other. On the other hand, participants 3, 4, 6, and 10 demonstrate no effect of set size but show a consistent effect of target presence. As we can see, the distributions for different set sizes largely overlap with each other in each target condition but differ considerably across target conditions. 

We refer to the sub-groups as the zero-effect and the target-effect groups, respectively. In the following sections, we examine the effectiveness of the different measures of distributional changes as estimates of experimental effects by their ability to portray the observed differences between the two sub-groups.

\subsection{Changes of RT distributions across Conditions}

Figure \ref{fig:six_distances} shows the distributional changes across all conditions measured in six different ways. The results are presented as distance maps by comparing RT distributions from all eight conditions to each other. In the maps, the first half of each axis corresponds to target-present trials and the second half to target-absent trials, and set sizes increase in ascending order within each half. In this setup, the changes in constant target conditions are presented in the bottom left and the top right quadrants; the effects of target presence in constant set sizes are presented in the top left and the bottom right quadrants.

The first two rows show RE and OI measures (Figs. a and b, respectively) that directly estimate the differences in distributions. As we can see, participants 3, 4, 6, and 10 (from the target-presence group) demonstrate maps that are very similar to each other. These four participants show nearly zero differences with setsize (bottom left and top right quadrants). On the other hand, they show large differences with the target condition (top left and bottom right quadrants). For the zero-effect group, we see the changes to be much more random compared to the patterns observed for the target-effect group. On average, the magnitudes of change were also quite low, which is discussed in the next section.

The third and fourth rows correspond to the changes in the parameters of Ex-G distributions, measured by ED and CS (Figs. c and d, respectively). The fifth and sixth rows correspond to the changes in the DDM parameters, again measured using ED and CS (Figs. e and f, respectively). These distance maps can be observed to show similar patterns to those in the RE and OI maps. For example, all six distances show similar maps for participants 3 and 10. However, these four maps also appear to be relatively noisier and do not show the subgroups as clearly as RE and OI maps do. Especially for participants 4 and 6, we see some disagreements among the six distances, but most of them reflect the same pattern of effects as captured by the RE and OI measures. Altogether, these results indicate that direct estimations of distributional changes (as with RE and OI) provide sensitive measures of the changes in RT distributions across conditions as estimates of the experimental effects.

\subsection{Average Experimental Effects on Individuals}

In this section, we examine the observations from the distance maps in the average effects across experimental conditions for each individual. The results are shown in Figure \ref{fig:effects_on_individuals}. In each subplot, we show the average set size vs target-presence effects as measured by each of the six distance measures. We marked the individuals from the zero-effect group with red cross signs and those from the target-effect group with green plus signs. 

A first point to note is that the effects occur at different scales of magnitude, as demonstrated by the scale differences in the X and Y axes. The set size effects are much lower than the target-presence effect, confirming our observations in previous sections.

Second, the two groups are well separated in the effects measured by RE and OI (Figs. a and b). The four participants from the target-effect group demonstrate relatively high effects of target conditions compared to the participants from the zero-effect group, whereas both groups show low setsize effects. In contrast, the changes in the Ex-G and the DDM parameters (measured using ED and CS) do not show as clear a divide as shown by the two direct measures of distributional change, with the ED measures slightly outperforming the CS measures.

Third, relative to the small scales of the observed effects, individuals demonstrate large differences in experimental effects among each other and even within each group. For all measures, the individuals span a considerable area in the 2D effect space, showing that the effects on individuals are much smaller compared to inter-individual variations of effects within the groups.

Finally, these variations also indicate the absence of an ``average" subject that represents the whole group well. The average participant is shown using black stars in Figure \ref{fig:effects_on_individuals}. Interestingly, for all measures, the effects on the average subject appear much lower compared to the individual-level effects. More importantly, their locations on these plots reinstate the perils of averaging. At best, the effects on the average participant may represent the majority of the individuals in the groups. At worst, the averages may fall in ``no man's land" and reflect no one in the group -- an outcome that appears true for most of the measures used in this study.

\section{Discussion}\label{sec:discussion}

Understanding individual-level effects is crucial in real-world contexts where group-level estimates may obscure fine-grained effects on individuals. In domains such as precision psychiatry, education, and user experience design, tailoring interventions based on identified individual characteristics can lead to considerably improved personalized solutions. However, progress in this direction has been hindered by the lack of simple and generalizable methods for investigating experimental effects at the individual level. 

Addressing this gap, we explore a distributional approach to examine experimental effects at the individual level. By analyzing RT distributions from 10 participants performing a visual search task across various set sizes and target-presence conditions, we uncovered significant individual differences in responses to the experimental manipulations.

Our visualization of the RT distributions revealed two distinct groups of participants. The first group (6 participants) exhibited no to minimal effects of the treatment conditions, suggesting consistency in their information processing strategies across conditions. In contrast, the second group demonstrated substantial effects of target presence, processing information for a longer duration when targets were absent. This divergence likely reflects differences in cognitive strategies. While one group may rely on similar processes regardless of condition, the other may perform exhaustive searches or additional verifications to confirm target absence. Additional data, such as from eye-tracking, could elucidate these behavioral patterns.

To quantify these effects, we employed two measures of distributional differences -- specifically, RE and OI -- and compared them against traditional models of RT distributions. Our results demonstrate that RE and OI captured intra-individual changes more clearly than the parametric models, despite the latter's ability to provide useful information about RT distributions. This result emphasizes the advantage of non-parametric measures for characterizing subtle distributional variations and augmenting domain-specific models as an additional layer of support and verification for the experimental effects on individuals.

Our results also illustrate how averaging across individuals could obscure the distinct strategies employed by participants. As we demonstrated, the averages often fail to represent any actual participant and lead to a "no man's land" scenario where group-level metrics may mislead interpretations. These findings highlight the perils of focusing exclusively on group means and reinforce the necessity to explore individual-level effects to uncover the nuanced dynamics of human information processing.

Finally, despite these insights, our study has several limitations. The small sample size of 10 participants and the focus on just two experimental effects restrict generalizability. Moreover, the lack of data beyond RT samples limits our ability to confirm the cognitive strategies underlying the observed RT differences. To overcome these limitations, future research could incorporate diverse multimodal data (e.g., eye-tracking, neuroimaging, phenotypes) from a larger number of participants to validate and extend our findings. Such efforts could establish distributional differences as a robust marker of behavioral change, as well as help develop individual-level models of information processing for personalized interventions.

\section{Summary and Conclusions}

In this work, we present a distributional spotlight to shed light on the experimental effects on individuals. Examining the effects on 10 individuals in a visual search task, we show that the distributional approach excels in capturing changes within individuals, as well as highlighting the individual differences across individuals. By revealing individual differences that group averages obscure, this approach offers a clearer view of the nuances of human behavior and its changes. While more tests are needed to establish the generalizability of this approach, our results underline the need to go beyond group means and into the world of individuals for precise estimations of experimental effects. Such precision may pave the way for refining cognitive models and tailoring interventions, advancing our theoretical understanding and enabling real-world applications such as precision mental health care.

\section*{Acknowledgments}
I sincerely thank Wayne D. Gray, Michael J. Schoelles, Christopher R. Sims, Sama Rahman, Ropa Denga, and Xue Zhang for their insightful feedback and brainstorming sessions, which greatly shaped this work. I am also grateful to Jeremy M. Wolfe, Evan M. Palmer, and Todd S. Horowitz for making their visual search dataset openly available and enabling this research. This study was conducted using personal time and resources, leveraging open-source data and tools, and using LLM-based AI agents for assistance in draft review and grammar checks. The codes and the results used in this work can be found here: \url{https://github.com/Roussel006/Individual-Differences-of-Choice-Reaction-Time-Distributions}.

\printbibliography

\end{document}